\documentclass[twocolumn]{aastex62}

\usepackage{CJK}

\graphicspath{{./}{figures/}}

\received{XXX}
\revised{YYY}
\accepted{ZZZ}
\submitjournal{ApJ}

\usepackage{graphicx}	
\usepackage{amsmath}	
\usepackage{amssymb}	
\usepackage[inline]{enumitem}
\usepackage{gensymb}
\usepackage{array}
\newcolumntype{P}[1]{>{\centering\arraybackslash}p{#1}}
\newcolumntype{M}[1]{>{\centering\arraybackslash}m{#1}}

\usepackage{color}

\newcommand{\pmn}{\texttt{PyMultiNest}}

\shorttitle{}
\shortauthors{}

\begin{document}
\begin{CJK*}{UTF8}{gbsn}

\title{Nodal Precession and Tidal Evolution of Two Hot-Jupiters: WASP-33 b and KELT-9 b}

\correspondingauthor{Alexander P. Stephan}
\email{stephan.98@osu.edu}

\author[0000-0001-8220-0548]{Alexander P. Stephan}
\affiliation{Department of Astronomy, The Ohio State University, 100 W 18th Ave, Columbus, OH 43210 USA}
\affiliation{Center for Cosmology and AstroParticle Physics, The Ohio State University, Columbus, OH 43210, USA}

\author[0000-0002-4361-8885]{Ji Wang (王吉)}
\affiliation{Department of Astronomy, The Ohio State University, 100 W 18th Ave, Columbus, OH 43210 USA}

\author[0000-0001-9207-0564]{P. Wilson Cauley}
\affiliation{Laboratory for Atmospheric and Space Physics, University of Colorado Boulder, Boulder, CO 80303, USA}

\author[0000-0003-0395-9869]{B. Scott Gaudi}
\affiliation{Department of Astronomy, The Ohio State University, Columbus, OH 43210, USA}

\author[0000-0002-0551-046X]{Ilya Ilyin}
\affiliation{Leibniz-Institut f{\"u}r Astrophysik Potsdam (AIP), An der Sternwarte 16, 14482 Potsdam, Germany}

\author[0000-0002-5099-8185]{Marshall C. Johnson}
\affiliation{Department of Astronomy, The Ohio State University, Columbus, OH 43210, USA}

\author[0000-0002-6192-6494]{Klaus G. Strassmeier}
\affiliation{Leibniz-Institut f{\"u}r Astrophysik Potsdam (AIP), An der Sternwarte 16, 14482 Potsdam, Germany}




\begin{abstract}

Hot Jupiters orbiting rapidly rotating stars on inclined orbits undergo tidally induced nodal precession measurable over several years of observations. The Hot Jupiters WASP-33 b and KELT-9 b are particularly interesting targets as they are among the hottest planets found to date, orbiting relatively massive stars. Here, we analyze archival and new data that span 11 and 5 years for WASP-33 b and KELT-9 b, respectively, in order to to model and improve upon their tidal precession parameters. Our work confirms the nodal precession for WASP-33 b and presents the first clear detection of the precession of KELT-9 b. We determine that WASP-33 and KELT-9 have gravitational quadrupole moments $(6.3^{+1.2}_{-0.8})\times10^{-5}$ and $(3.26^{+0.93}_{-0.80})\times10^{-4}$, respectively.
We estimate the planets' precession periods to be $1460^{+170}_{-130}$ years and $890^{+200}_{-140}$ years, respectively, and that they will cease to transit their host stars around the years $2090^{+17}_{-10}$~CE and $2074^{+12}_{-10}$~CE, respectively. Additionally, we investigate both planets' tidal and orbital evolution, suggesting that a high-eccentricity tidal migration scenario is possible to produce both system architectures and that they will most likely not be engulfed by their hosts before the end of their main sequence lifetimes.

\end{abstract}


\section{Introduction}\label{sec:intro}
 
Over recent decades the number of discovered exoplanets has steadily increased into the thousands, revealing a large variety of exoplanet system orbital architectures around stars of any life stage covering a large range of stellar masses and life stages \citep[e.g.,][]{Howard+2012,Charpinet+2011,Barnes+2013,Gettel+2012,Nowak+2013,Reffert+2015,Niedzielski+2015,Niedzielski+2016}. Of peculiar interest among these exoplanetary systems have been Hot Jupiters (HJs), which orbit their stars with periods of 10 days or shorter, as such planets do not exist in our own solar system. Additionally, many discovered HJs have strong spin-orbit misalignments with their host stars, often even being in retrogarde orbits \citep[e.g.,][]{Albrecht+2012,WinnFab2015}. This, together with planet formation models, supports the idea that HJs are not born at their current orbital configuration, but have migrated post-formation by some dynamical mechanism \citep[e.g.,][]{Nag+08,Naoz11,Naoz+11sec,Naoz+12bin}.

Regardless of the formation or migration mechanisms, spin-orbit misaligned HJs present a unique observational opportunity to measure tidally induced nodal precession of an HJ's orbit, if the host star is oblate due to rapid rotation. As colder, more convective stars tend to ``spin down'' rapidly due to magnetic braking as they age \citep[e.g.,][]{Kawaler1988}, hotter ($T_{eff}>7000$~K), stars with radiative envelopes can stay fast-spinning for most of their main-sequence (MS) life time \citep[e.g.,][]{Meynet+2011}. Therefore, misaligned HJs around hot stars are ideal targets for measuring nodal precession. 

Here, we present our observations and analysis of the nodal precession and tidal dynamics of two HJs orbiting two hot, rapidly spinning stars, WASP-33 b and KELT-9 b. These exoplanets are among the hottest exoplanets that have been discovered over recent decades, orbiting stars with effective temperatures above $7400$~K and $10,000$~K, respectively \citep[e.g.,][]{Cameron+2010,Gaudi+2017}. Both planets have been extensively studied by previous works, investigating among other factors their orbits and atmospheres \citep[e.g.,][]{Cameron+2010,vonEssen+2014,johnson15,Gaudi+2017,Ahlers+2020,watanabe20,Borsa+2021,Asnodkar+2022b,Asnodkar+2022a}, and are known to have large spin-orbit misalignments. In this work we obtain new data to extend the timeline of nodal precession measurements by several years for both planets compared to previous works (see Sections \ref{sec:obs} and \ref{sec:modelDT}), allowing us to derive more accurate tidal parameters for both systems (see Section \ref{sec:prec}). Additionally, we estimate the past tidal evolution of both planets, allowing us to set limits to the formation and migration mechanisms that brought them to their current configurations(see Section \ref{sec:disc}).

\section{Observations}\label{sec:obs}
\subsection{WASP-33 b}

We use archival data for WASP-33 b that were obtained by Harlan J. Smith Telescope (HJST) with Robert G. Tull Coude Spectrograph~\citep{Tull1995} at McDonald Observatory on UT 2008 November 12th~\citep{Cameron2010} and UT 2014 October 4th~\citep{johnson15}. 

In addition, we include new data from HJST that were obtained on UT 2016 December 11 and data from the PEPSI \citep{strassmeier15} spectrograph on the Large Binocular Telescope (LBT) on UT 2019 November 17~\citep{Cauley2021}. 

There are also archival data from the Subaru telescope. However, we do not have access to the Subaru data that were obtained on UT 2011 October 19~\citep{watanabe20}. Because the Subaru observation was bracketed by the two ends of the time baseline from 2008 to 2019, the Subaru data are non-essential in the subsequent analyses. For a similar reason, we do not include archival data from HARPS-N that span from 2016 to 2019~\citep{Borsa+2021}.   

\subsection{KELT-9 b}

We use archival data in the discovery paper~\citep{Gaudi+2017}. These data were obtained by the Tillinghast Reflector Echelle Spectrograph (TRES) on the 1.5 m telescope at the Fred Lawrence Whipple Observatory on UT 2014 November 14, 2015 November 06, and 2016 June 12.

We also include archival data that were obtained since the discovery. The data include two transits observed with the HARPS-N spectrograph mounted on the Telescopio Nazionale Galileo (TNG) in La Palma, Spain on UT 2017 Aug 01 and 2018 Jul 21~\citep{Hoeijmakers2019}, and one transit observed with PEPSI at LBT on UT 2018 Jul 03~\citep{Cauley2019}. 

In addition, we add a new data set from PEPSI at LBT that were observed on UT 2019 Jun 22, therefore extending the time baseline to $\sim$5 years from 2014 to 2019. We note that CARMENES~\citep{Quirrenbach2018} has also observed KELT-9 b~\citep{Yan2019}. The observation was made on UT 2017 Jan 5. We do not include this data point because it falls in the time baseline and does not add significant constraint on the nodal precession measurement. 

\section{Modeling the Doppler Tomography (DT) Signal}\label{sec:model}

Both WASP-33 b and KELT-9 b have been studied with DT~\citep{Gaudi+2017, johnson15, watanabe20}. With the new data sets that have been taken in more recent epochs, we perform a uniform analysis to model the DT signal for the two planets. The uniform analysis will eliminate any systematics introduced by using different models. 

\begin{figure*}[htbp]
   \centering
    \includegraphics[width=0.6\linewidth]{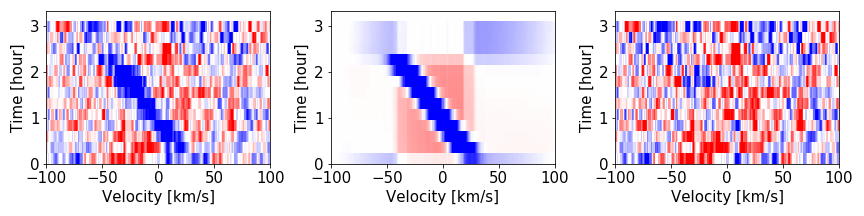}\\
    \includegraphics[width=0.6\linewidth]{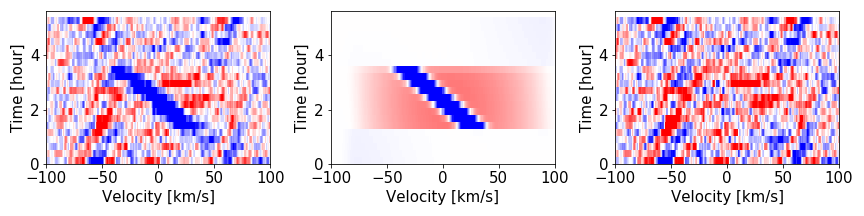}\\
    \includegraphics[width=0.6\linewidth]{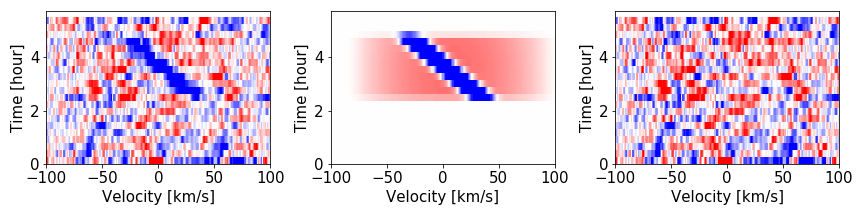}\\
    \includegraphics[width=0.6\linewidth]{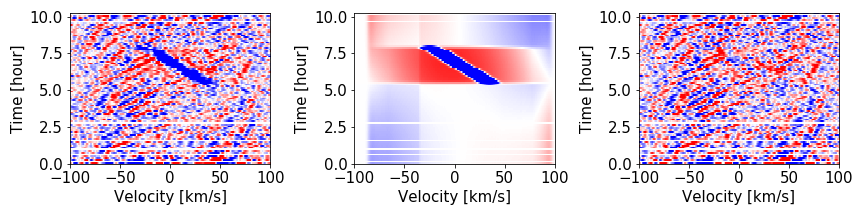}\\

    \figcaption{From top to bottom are our modeling results for WASP-33 b data in 2008 (TULL), 2014 (TULL), 2016 (TULL), and 2019 (PEPSI). {\bf{Left}}: Fourier-filtered residual map after subtracting the median line profile.
   The planet ``Doppler shadow" is the diagonal blue track running from the bottom-right to the
   top-left. Occasional gaps are due to interpolation onto a time array with a fixed interval. {\bf{Middle}}: Modeled DT signal. The overall gradient along the time axis is due to the shift of stellar radial velocity. The enhanced red background during transit is due to the fact that we set the line profile normalization to unity for all line profiles.  {\bf{Right}}: difference of the two maps on the left and middle panel. 
\label{fig:wasp33b_dtmap}}

\end{figure*}

\subsection{Extracting Line Profiles (LPs)}

We follow the least-square deconvolution~\citep[LSD, ][]{Kochukhov2010, Pai2021} method to extract line profiles (LPs). The LSD method requires a stellar template spectrum, for which we use the PHOENIX BT-Settl model~\citep[][and referneces therein]{Allard2012, Allard2013}. We do not interpolate between stellar parameters. Instead, we use the spectra whose effective temperature, surface gravity, and metallicity are the closest to the reported values for WASP-33 b and KELT-9 b, i.e., T$_{\rm{eff}}$ = 7400 K, log(g) = 4.0, and solar metallicity for WASP-33 b, and T$_{\rm{eff}}$ = 10200 K, log(g) = 4.0, and solar metallicity for KELT-9 b. 

The LSD method also requires a scalar value for the regularization matrix. The scalar controls the smoothness of the extracted LPs. We experiment with a wide range of possible values that span 4 orders of magnitude and choose a scalar that retain the DT signal and planet absorption signal (if visible) while removing the stochastic noise induced by the inversion process of LSD.  

\subsection{Removing Stellar Pulsation}
WASP-33 is a $\delta$-Scuti variable star that shows a very strong stellar pulsation signal. The stellar pulsation is so strong that it overwhelms the DT signal. As detailed in~\citet{Cauley2021}, we apply a customized Fourier filter to remove the pulsation signal. Because the pulsation signal and the DT signal are almost orthogonal 
to each other, the two signals can be easily separated in Fourier space. A similar
strategy was adopted in previous works~\citep{johnson15,watanabe20}.

We do not attempt to remove the stellar pulsation signal for KELT-9 b although it was suggested in~\citet{Hoeijmakers2019}. This is because the pulsation signal does not pose a challenge for us in the modeling of the DT signal for KELT-9 b.

\subsection{Modeling DT}
\label{sec:modelDT}
We adopt an analytical framework to model the DT signal as described in~\citet{Cauley2021}. In short, the DT signal is a Gaussian perturbation of the stellar line
profile~\citep{Hirano2011}. Therefore, the DT signal can be analytically calculated and the details are described in~\citet{Wang2018}. The analytical framework can be applied to modeling not only the DT signal~\citep[e.g., WASP-33 b, ][]{Cauley2021}, but also (1) the planet absorption signal for high-resolution spectroscopy (as shown for KELT-9 b); and (2) photometric and spectroscopic observations of eclipsing binaries~\citep[e.g., EPIC 203868608, ][]{Wang2018}. 

\begin{figure*}[htbp]
   \centering
    \includegraphics[width=0.6\linewidth]{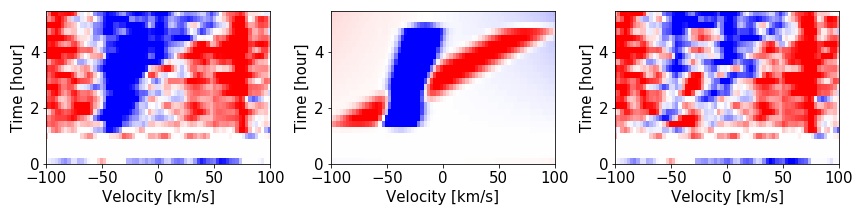}\\
    \includegraphics[width=0.6\linewidth]{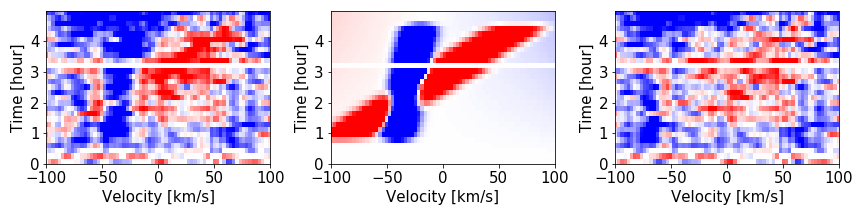}\\
    \includegraphics[width=0.6\linewidth]{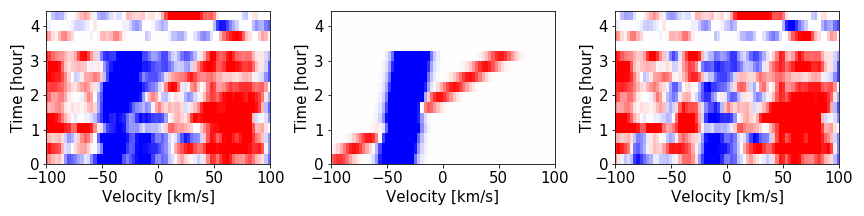}\\
    \includegraphics[width=0.6\linewidth]{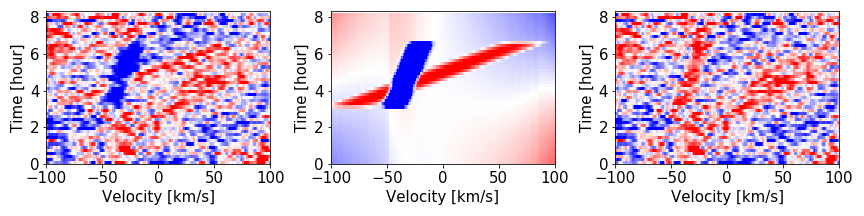}\\
    \includegraphics[width=0.6\linewidth]{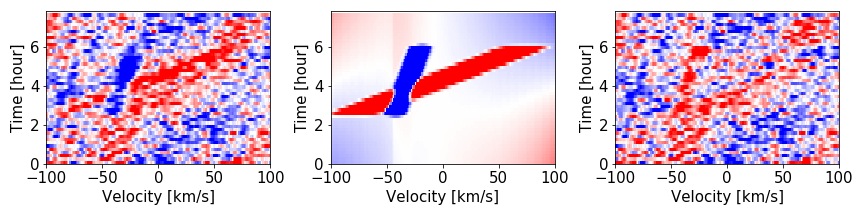}\\
    \includegraphics[width=0.6\linewidth]{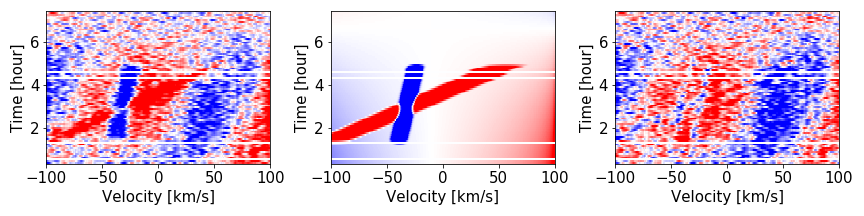}\\
    \includegraphics[width=0.6\linewidth]{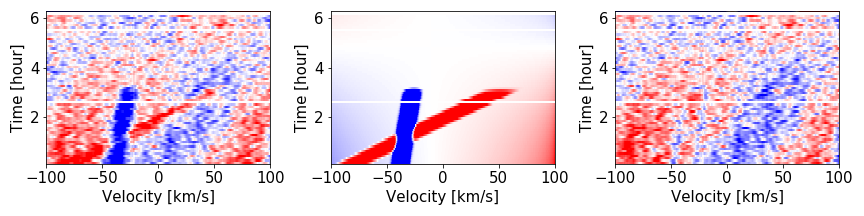}\\
    \figcaption{From top to bottom are our modeling results for KELT-9 b data in 2014 (TRES), 2015 (TRES), 2016 (TRES), 2017 (HARPS-N), 2018 (HARPS-N), 2018 (PEPSI), and 2019 (PEPSI). {\bf{Left}}: Residual map after subtracting the median line profile.
   The planet ``Doppler shadow" is the vertical blue track due to the nearly polar orbit of KELT-9 b. The planet absorption signal is the diagonal red track that runs from bottom left to top right. The planet signal is more apparent in 2017-2019 data taken from HARPS-N and PEPSI due to a higher signal-to-noise ratio. Occasional gaps are due to interpolation onto a time array with a fixed interval. {\bf{Middle}}: Modeled DT  and planet absorption signal.  {\bf{Right}}: difference of the two maps on the left and middle panel. 
\label{fig:kelt9b_dtmap}}

\end{figure*}

For WASP-33 b, the modeling parameters for the DT signal are the impact parameter, projected spin-orbit alignment angle, projected rotational velocity, quadratic limb darkening parameters, planet-star radius ratio, systemic velocity, and the mid-transit time. For KELT-9 b, the planet absorption signal during the transit is clearly visible, so we add the following parameters: planet-star mass ratio, planet absorption depth and width, and a net velocity shift in the planet rest frame. Fig. \ref{fig:wasp33b_dtmap} and Fig. \ref{fig:kelt9b_dtmap} show the modeling results for WASP-33 b and KELT-9 b. 

\subsection{Sampling Posteriors}\label{subsec:post}

We used \pmn~\citep{Buchner2014} for posterior sampling in a Bayesian framework. We fit each data set independently and obtain posterior samples for all parameters as described in \S \ref{sec:modelDT}. In this work, we focus on measuring the nodal precession as manifested by the change of projected orbital obliquity ($\lambda$) and impact parameter ($b$). Priors are uniform distributions between -180 degree and 180 degree for $\lambda$ and between -1 and 1 for $b$. The likelihood function is $\exp[{-(\mathcal{D}-\mathcal{M})^2/\mathcal{E}^2}]$, where $\mathcal{D}$ is data, $\mathcal{M}$ is model, and $\mathcal{E}$ is the error term, which we use as the standard deviation of the difference between the LPs and the median LP over time. The posterior distributions of $\lambda$ and $b$ are summarized in Table \ref{tab:b_lambda}.

\subsection{Estimating Systematics}\label{subsec:system}
We eliminate systematics by uniformly analyzing all data sets using the analytical modeling framework as described in previous sections. However, our model has limitations in describing physical process of a planet occulting the stellar disk. For example, the perturbation of the planet shadow to the LP may not be sufficiently described by our analytical model, which is a convolution of a rotational-broadening kernel and an Gaussian instrument profile. This could explain some of the residuals that we see in the high signal-to-noise ratio (SNR) case, e.g., the lower four panels in Fig. \ref{fig:kelt9b_dtmap} for data from HARPS-N and PEPSI. For the low-SNR cases, e.g., 2014-2-16 data from TRES for KELT-9 b (top three panels in Fig. \ref{fig:kelt9b_dtmap}), we also notice structured residual patterns after subtracting the model from the data. 

To estimate the systematics in the nodal precession measurement, we take the following two approaches. First, we have two independent data sets for KELT-9 b that are taken close in time: HARPS-N data on 2018 Jul 20 and PEPSI data on 2018 Jul 03. The difference of results for $\lambda$ and $b$ between these two data sets can be used to estimate the systematics, under the assumption that the nodal precession is smooth over time scale of a year. The difference of $\lambda$ and $b$ between the two data sets is 0.73 degree and 0.01, respectively. This is at least $\sim$4 and $\sim$6 times larger than the uncertainty based on the posterior from the best fit models for $\lambda$ and $b$ respectively (See Table \ref{tab:b_lambda}). 

The second approach we take is to slightly change our data analyses procedure and check how much the results vary accordingly. For example, instead of subtracting the out-of-transit median LP, we subtract the overall median LP. We find that the change in the analyses procedure can result in the change of $\lambda$ and $b$ by $\sim$2-3 times of the reported uncertainty.  

Given the above exercises in understanding the systematics in the analyses, we inflate the measurement uncertainties for $\lambda$ and $b$ to 0.75 degree and 0.01 for cases which return lower than these uncertainties based on posterior distributions, in order to account for unknown sources of systematics.

\section{Modeling Planetary Precession}\label{sec:prec}

\begin{figure*}[htbp]
   \centering
    \includegraphics[width=0.4\linewidth]{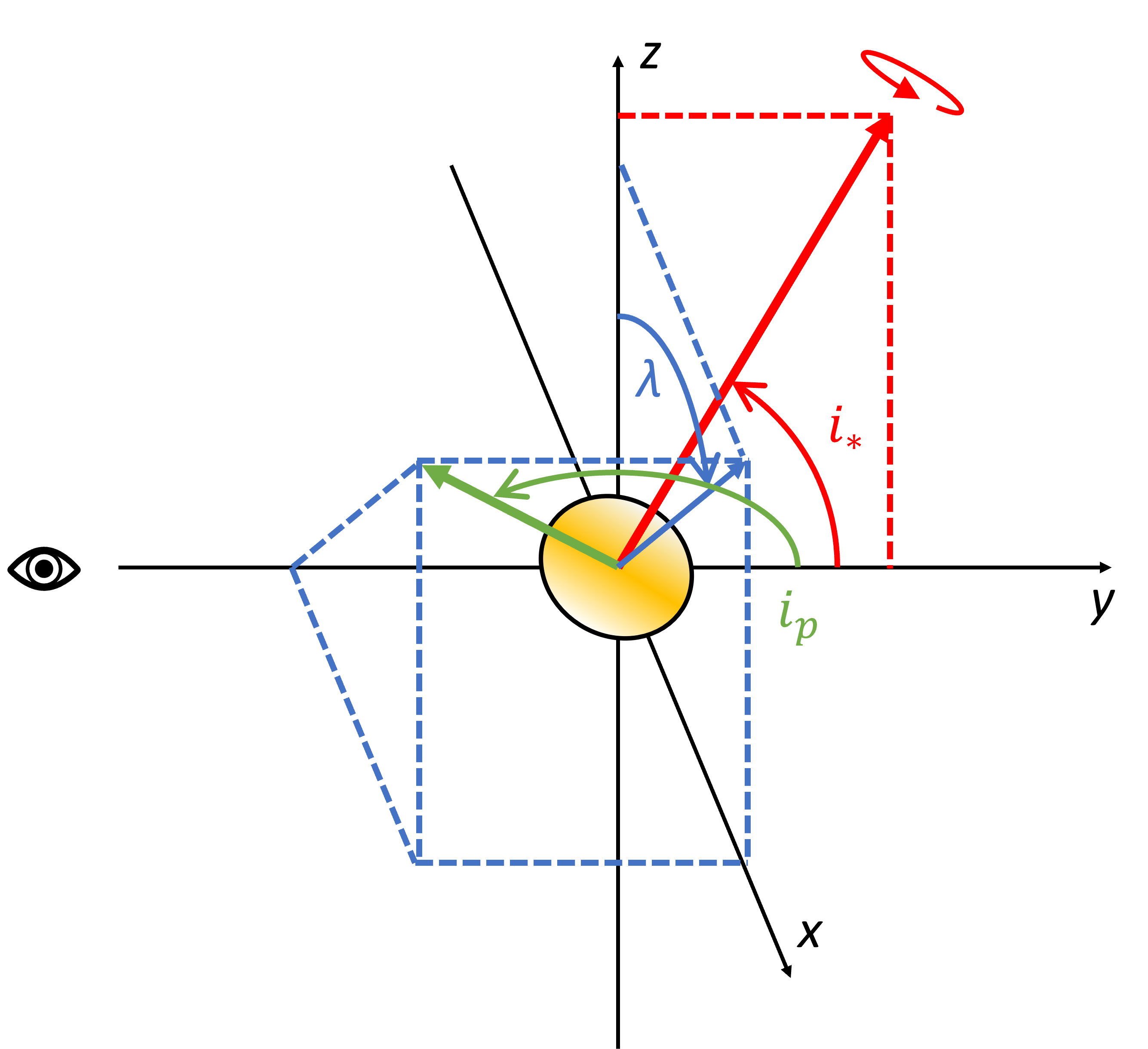}
    \includegraphics[width=0.4\linewidth]{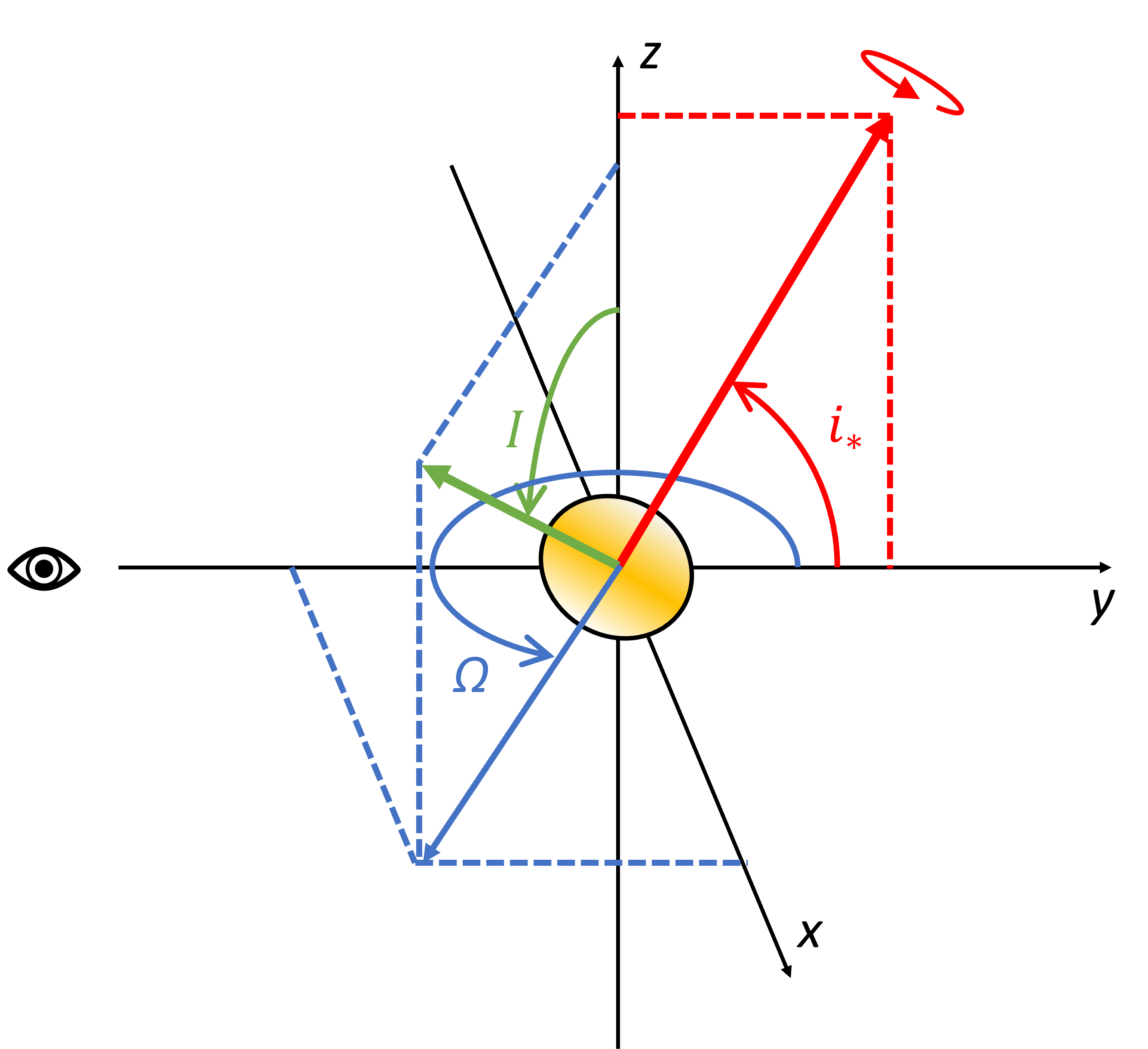}

    \figcaption{Geometry of the orbital elements and observables as defined in this study. The observer's line of sight is defined as the $y$-axis, with the sky defined along the $xz$ plane. The stellar spin axis is defined to be on the $yz$ plane (see solid red arrow), forming an angle $i_*$ with the line of sight. If the star is viewed equator-on, $i_*=90$~deg. The orbital angular momentum vector of the planet (see green solid arrow) forms an angle $i_p$ with the line of sight and an angle $I$ with the $z$-axis ($I$ is also the inclination between the orbital plane and the $xy$ plane). The projection of the orbital angular momentum vector onto the $xz$ plane defines the planet's projected obliquity versus the $z$ axis, $\lambda$, while the projection of the vector onto the $xy$ plane defines the ascending node versus the line of sight, $\Omega$. See also \citet{watanabe20}, Figure 4., on which this graphic is based.
\label{fig:orientation}}
\end{figure*}

We model the observed changes of the projected obliquity $\lambda$ and impact parameter $b$ over time, assuming precession of the nodes due to tidal interactions between the planets and their host stars. Our analysis is similar to that discussed in previous works \citep[e.g.,][]{watanabe20,Borsa+2021}. We translate the observed values of $\lambda$ and $b$ into values for the ascending node versus the line-of-sight, $\Omega$, and the inclination of the planetary orbital plane versus the plane of the sky, $I$, via the equations
\begin{equation}\label{eq:tanOmega}
    \tan \Omega = -\sin \lambda \tan i_p
\end{equation} 
and
\begin{equation}\label{eq:cosI}
    \cos I = \cos \lambda \sin i_p \ ,
\end{equation} 
where $i_p$ is the angle between the observer's line-of-sight and the planet's angular momentum. The latter can be derived given the impact parameter $b$ and the ratio of the planetary orbit's semi-major axis to the star's radius, $(a/R_*)$, via
\begin{equation}\label{eq:b_i_p}
    b = (a/R_*) \cos i_p \ .
\end{equation} 
The values for these parameters for the different observation dates are shown in Table \ref{tab:b_lambda}. The geometry of these various angles is shown in Fig. \ref{fig:orientation}.

The long-term precession of the longitude of the ascending node, $\Omega$, and inclination, $I$, of any planet with significantly smaller orbital angular momentum than its host star's stellar rotational angular momentum can be described by the equations \citep[e.g.,][]{Iorio2016} \begin{equation}\label{eq:Omega_dot}
\begin{aligned}
    \dot{\Omega} = -\frac{3 \pi J_2 R_*^2}{2 P a^2} & \{ 2 \sin{i_*} \cos{i_*} \cos{2 I} \csc{I} \cos{\Omega} \\
    &- \cos{I}\left( 1 - 3 \sin^2{i_*} + \cos^2{i_*} \cos{2 \Omega} \right) \},
\end{aligned}
\end{equation}

 and \begin{equation}\label{eq:I_dot}
\begin{aligned}
    \dot{I} = -\frac{3 \pi J_2 R_*^2}{P a^2} \cos{i_*}\sin{\Omega} \{& \sin{i_*}\cos{I} \\ &- \cos{i_*}\sin{I}\cos{\Omega} \},
\end{aligned}
\end{equation} with $i _*$ being the angle of the star's rotation axis to the line-of-sight of the observer, $P$ being the planet's orbital period, $a$ being the planet's orbital semi-major axis, $R_*$ being the stellar radius, and $J_2$ being the star's quadrupole gravitational moment. The values of most of these parameters are taken from the previous literature and are listed in Table \ref{tab:literature}.

We use Equations \ref{eq:Omega_dot} and \ref{eq:I_dot} to calculate the evolution of $\Omega$ and $I$ over time and fit them to our observed data in Table \ref{tab:b_lambda} via maximum likelihood estimation, with $i_*$, $J_2$, $\Omega_0$, and $I_0$ as free parameters to be determined by the best-fit solution. To determine the confidence intervals for these parameters we use the MCMC code {\tt emcee} \citep{Foreman-Mackey+2013}.

\begin{table*}
	\centering
	\caption{Measured values for $\lambda$ and $b$ and derived values for $i_p$, $\Omega$, and $I$ for WASP-33 b and KELT-9 b. Given are the 68th-percentile confidence intervals for each parameter determined from the posterior distributions discussed in Section \ref{subsec:post}. As mentioned in Section \ref{subsec:post}, we applied two analysis procedures to the KELT-9 b data. As explained in Section \ref{subsec:system}, for our precession model we inflated the uncertainties of $\lambda$ and $b$ to minimum values of $0.75\degree$ and $0.01$, respectively, where necessary.}
	\label{tab:b_lambda}
	\begin{tabular}{m{0.1\textwidth}|M{0.1\textwidth}M{0.1\textwidth}M{0.1\textwidth}M{0.1\textwidth}} 
	    \multicolumn{5}{c}{WASP-33 b} \\
		\hline
		Source & HJST & HJST & HJST & PEPSI \\
		Mid-transit Time [UT] & 2008.11.12\newline10:02 & 2014.10.04\newline06:29 & 2016.12.11\newline06:52 & 2019.11.17\newline07:59 \\
		\hline
		$\lambda$ [$\degree$] & $-109.82^{+0.54}_{-0.55}$ & $-110.27^{+0.40}_{-0.41}$ & $-110.29^{+0.55}_{-0.60}$ & $-109.97^{+0.41}_{-0.41}$ \\
		$b$ & 
		$0.1502^{+0.0199}_{-0.0075}$ & $0.0821^{+0.0090}_{-0.0059}$ & $0.0548^{+0.0084}_{-0.0143}$ & $-0.0074^{+0.0077}_{-0.0080}$\\
		$i_p$ [$\degree$] & 
		$87.67^{+0.12}_{-0.31}$  & $88.72^{+0.09}_{-0.14}$ & $89.15^{+0.22}_{-0.13}$ & $90.11^{+0.12}_{-0.12}$\\
		$\Omega$ [$\degree$] & $87.52^{+0.13}_{-0.33}$  & $88.64^{+0.10}_{-0.15}$ & $89.09^{+0.23}_{-0.14}$ & $90.12^{+0.13}_{-0.13}$\\
		$I$ [$\degree$] & 
		$109.80^{+0.55}_{-0.54}$ & $110.27^{+0.41}_{-0.40}$ & $110.28^{+0.60}_{-0.55}$ & $109.97^{+0.41}_{-0.41}$\\
		\hline
	\end{tabular}

	\centering
    \vspace*{0.5 cm}

	\begin{tabular}{m{0.1\textwidth}|M{0.1\textwidth}M{0.1\textwidth}M{0.1\textwidth}M{0.1\textwidth}M{0.1\textwidth}M{0.1\textwidth}M{0.1\textwidth}M{0.1\textwidth}} 
	    \multicolumn{8}{c}{KELT-9 b (1st analysis procedure)} \\
		\hline
		Source & TRES & TRES & TRES & HARPS-N & PEPSI & HARPS-N & PEPSI \\
		Mid-transit Time [UT] & 2014.11.14\newline05:09 & 2015.11.06\newline03:58 & 2016.06.12\newline08:55 & 2017.08.01\newline02:04 & 2018.07.03\newline07:14 & 2018.07.21\newline01:48 & 2019.06.22\newline06:57 \\
		\hline
		$\lambda$ [$\degree$] & $-86.23^{+0.26}_{-0.23}$ & $-85.06^{+0.70}_{-0.84}$ & $-87.36^{+0.41}_{-0.34}$ & $-84.03^{+0.31}_{-0.33}$ & $-85.67^{+0.16}_{-0.17}$ & $-84.78^{+0.37}_{-0.38}$ & $-87.34^{+0.07}_{-0.08}$ \\
		$b$ & 
		$0.1176^{+0.0057}_{-0.0037}$ &
        $0.1297^{+0.0085}_{-0.0059}$ &
        $0.1766^{+0.0047}_{-0.0052}$ &
        $0.1912^{+0.0039}_{-0.0033}$ &
        $0.1873^{+0.0009}_{-0.0010}$ &
        $0.1954^{+0.0038}_{-0.0033}$ &
        $0.2004^{+0.0006}_{-0.0045}$ \\
		$i_p$ [$\degree$] & 
		$87.86^{+0.07}_{-0.10}$ &
        $87.64^{+0.11}_{-0.15}$ &
        $86.79^{+0.10}_{-0.09}$ &
        $86.52^{+0.06}_{-0.07}$ &
        $86.59^{+0.02}_{-0.02}$ &
        $86.45^{+0.06}_{-0.07}$ &
        $86.36^{+0.08}_{-0.01}$ \\
		$\Omega$ [$\degree$] & 
		$87.86^{+0.07}_{-0.10}$ &
        $87.63^{+0.11}_{-0.16}$ &
        $86.78^{+0.10}_{-0.09}$ &
        $86.50^{+0.06}_{-0.07}$ &
        $86.59^{+0.02}_{-0.02}$ &
        $86.43^{+0.06}_{-0.07}$ &
        $86.35^{+0.08}_{-0.01}$ \\
		$I$ [$\degree$] & 
		$86.23^{+0.23}_{-0.26}$ &
        $85.06^{+0.84}_{-0.70}$ &
        $87.37^{+0.34}_{-0.41}$ &
        $84.04^{+0.33}_{-0.31}$ &
        $85.68^{+0.17}_{-0.16}$ &
        $84.79^{+0.38}_{-0.37}$ &
        $87.34^{+0.08}_{-0.07}$ \\
		\hline
		\multicolumn{8}{c}{KELT-9 b (2nd analysis procedure)} \\
		\hline
		$\lambda$ [$\degree$] & $-85.20^{+0.26}_{-0.29}$ &
        $-85.26^{+0.57}_{-0.72}$ &
        $-86.69^{+0.43}_{-0.47}$ &
        $-84.58^{+0.30}_{-0.32}$ &
        $-85.91^{+0.09}_{-0.09}$ &
        $-85.18^{+0.33}_{-0.33}$ &
        $-86.53^{+0.08}_{-0.09}$  \\
		$b$ & 
		$0.1171^{+0.0056}_{-0.0037}$ &
        $0.1363^{+0.0073}_{-0.0054}$ &
        $0.1725^{+0.0042}_{-0.0051}$ &
        $0.1905^{+0.0036}_{-0.0029}$ &
        $0.1884^{+0.0007}_{-0.0014}$ &
        $0.1994^{+0.0031}_{-0.0031}$ &
        $0.2016^{+0.0006}_{-0.0010}$ \\
		$i_p$ [$\degree$] & 
		$87.87^{+0.07}_{-0.10}$ &
        $87.52^{+0.10}_{-0.13}$ &
        $86.86^{+0.09}_{-0.08}$ &
        $86.54^{+0.05}_{-0.07}$ &
        $86.57^{+0.02}_{-0.01}$ &
        $86.37^{+0.06}_{-0.06}$ &
        $86.33^{+0.02}_{-0.01}$ \\
		$\Omega$ [$\degree$] & 
		$87.86^{+0.07}_{-0.10}$ &
        $87.51^{+0.10}_{-0.13}$ &
        $86.86^{+0.09}_{-0.08}$ &
        $86.52^{+0.05}_{-0.07}$ &
        $86.57^{+0.02}_{-0.01}$ &
        $86.36^{+0.06}_{-0.06}$ &
        $86.33^{+0.02}_{-0.01}$ \\
		$I$ [$\degree$] & 
		$85.20^{+0.29}_{-0.26}$ &
        $85.27^{+0.72}_{-0.57}$ &
        $86.69^{+0.47}_{-0.43}$ &
        $84.59^{+0.32}_{-0.30}$ &
        $85.92^{+0.09}_{-0.09}$ &
        $85.19^{+0.33}_{-0.33}$ &
        $86.54^{+0.09}_{-0.08}$ \\
		\hline
		
	\end{tabular}
\end{table*}

\begin{table*}
	\centering
	\caption{Stellar and Planetary Parameters of WASP-33 and KELT-9 from the Literature used for the Nodal Precession and Tidal Evolution Models.}
	\label{tab:literature}
	\begin{tabular}{l|cc} 
	    \multicolumn{3}{c}{WASP-33} \\
		\hline
		Parameter & Value & Reference \\
		\hline
		$V\sin{i_*}$ (km/sec) & $86.63^{+0.37}_{-0.32}$  & \citet{johnson15} \\
		$a/R_s$ & $3.69\pm0.01$  & \citet{Cameron2010} \\
		$R_p/R_s$ & $0.1143\pm0.0002$ & \citet{Cameron2010} \\
		$R_s$ ($R_\odot$) & $1.444\pm0.034$  & \citet{Cameron2010}  \\
		$M_p$ ($M_{Jup}$) & $2.81\pm0.53$ & \citet{vonEssen+2014} \\
		$M_s$ ($M_\odot$) & $1.495\pm0.031$ & \citet{Cameron2010}   \\
		$P$ (days) & $1.2198675\pm0.0000011$ & \citet{vonEssen+2014} \\
		\hline
		\multicolumn{3}{c}{KELT-9} \\
		\hline
		Parameter & Value & Reference \\
		\hline
		$V\sin{i_*}$ (km/sec) & $111.4 \pm 1.3$  & \citet{Gaudi+2017} \\
		$a$ (AU) & $0.03462^{+0.00110}_{-0.00093}$ & \citet{Gaudi+2017} \\
		$R_p$ ($R_{Jup}$) & $1.891^{+0.061}_{-0.053}$ & \citet{Gaudi+2017} \\
		$R_s$ ($R_\odot$)& $2.362^{+0.075}_{-0.063}$ & \citet{Gaudi+2017} \\
		$M_p$ ($M_{Jup}$) & $2.88\pm0.84$ & \citet{Gaudi+2017} \\
		$M_s$ ($M_\odot$)& $2.52^{+0.25}_{-0.20}$ & \citet{Gaudi+2017} \\
		$P$ (days) & $1.4811235\pm0.0000011$ & \citet{Gaudi+2017} \\
		\hline
	\end{tabular}

\end{table*}

\subsection{WASP-33 b}\label{subsec:prec_WASP33}

\begin{figure}
    \hspace{0.0\linewidth}
    \includegraphics[width=\linewidth]{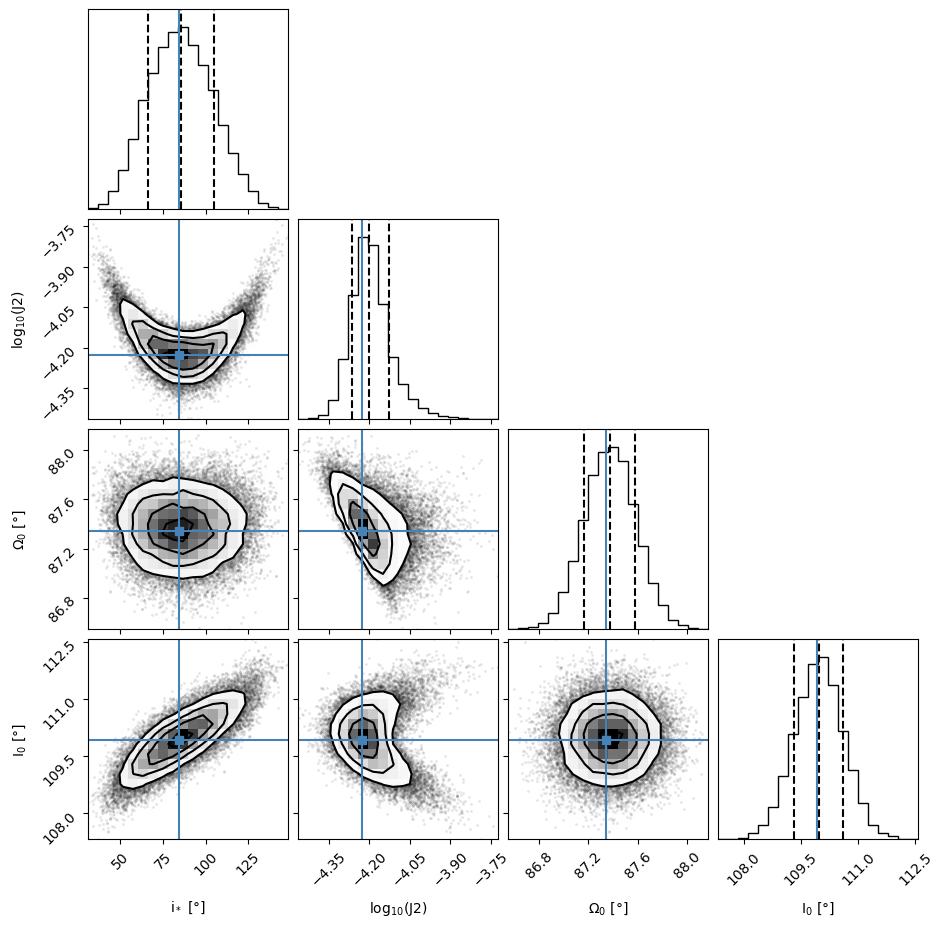}
    \figcaption{Corner plot of MCMC-determined values for $i_*$, $log_{10}(J_2)$, $\Omega_0$, and $I_0$ of WASP-33 b. The black, dashed lines mark the 1-$\sigma$ confidence interval for the values of each parameter, while the blue lines mark the parameter values of the best-fit model solution. See Table \ref{tab:results} for parameter values.
\label{fig:wasp33b_corner}}
\end{figure}

We apply the model discussed in the previous section to the WASP-33 b data, as shown in Table \ref{tab:b_lambda}, with inflated uncertainties as explained in Section \ref{subsec:system}. We use the earliest measured values and uncertainties for $\Omega$ and $I$ as priors for $\Omega_{0}$ and $I_{0}$. From the posterior we determined 1-$\sigma$ confidence intervals (determined by the 16th, 50th, and 84th percentiles) and best-fit model values of $J_2$, $i_*$, $\Omega_{0}$, and $I_{0}$, which we list in Table \ref{tab:results} (see Figure \ref{fig:wasp33b_corner} for a visualization of the probability distributions).

\begin{figure}
    \hspace{0.0\linewidth}
    \includegraphics[width=\linewidth]{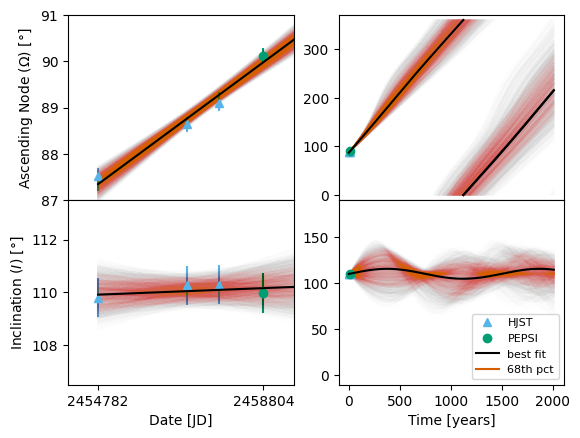}\\
    \includegraphics[width=\linewidth]{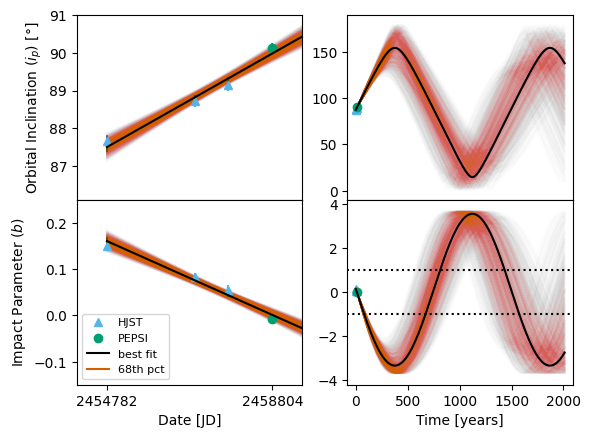}
    \figcaption{Evolution of the ascending node $\Omega$ (first row), inclination $I$ (second row), the planet's orbital inclination versus the observer's line-of-sight $i_p$ (third row), and impact parameter $b$ (fourth row) over time for WASP-33 b. The colored symbols and error bars mark the measured values and uncertainties of the parameters, with light blue triangles showing the HJST data and green dots showing PEPSI data, while the black line shows the best-fit model solution based on the data. The red lines show $2000$ example model solutions from our MCMC with model likelihoods within the 1-$\sigma$ confidence level of the best-fit model. The left frames focus on the time span of the measurements, while the right frames extrapolate the model for the next $2000$ years. The precession period for $\Omega$ is approximately $1460^{+170}_{-130}$ years. The value of $b$ must be between $-1$ and $1$ for observable transits to occur (see black dotted lines). Based on our model estimate, WASP-33 b will cease to transit around the year $2090_{-10}^{+17}$~CE.
\label{fig:wasp33b_precession}}
\end{figure}

These values translate to the precession model shown in Figure \ref{fig:wasp33b_precession}, with a precession period of roughly $1460^{+170}_{-130}$ years for the ascending node. In Figure \ref{fig:wasp33b_precession} we show the evolution of the ascending node $\Omega$, inclination of the orbital plane to the sky $I$, inclination of the planet's orbit to the line-of-sight $i_p$, and impact parameter $b$, respectively. Note that the planet can only transit in respect to an Earth-bound observer if $b$ is between $-1$ and $1$ (excluding ``grazing'' transits). Based on our model estimates, WASP-33 b's impact parameter will dip below $-1$ around the year $2090_{-10}^{+17}$ and cease to transit for approximately $500\pm200$ years (see last frame of Figure \ref{fig:wasp33b_precession}). Overall, WASP-33 b only transits for about $20~\%$ of its precession period.

\subsection{KELT-9 b}\label{subsec:prec_KELT9}

\begin{figure}
    \hspace{0.0\linewidth}
    \includegraphics[width=\linewidth]{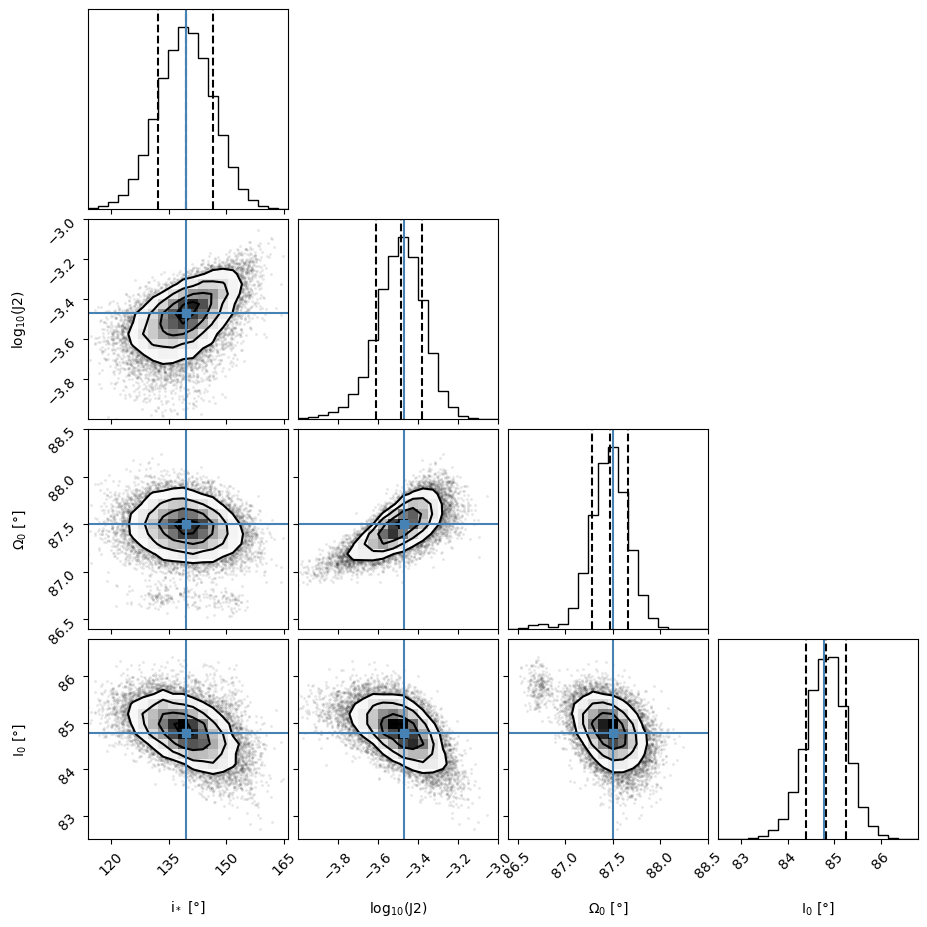}
    \figcaption{Corner plot of MCMC-determined values for $i_*$, $log_{10}(J_2)$, $\Omega_0$, and $I_0$ of KELT-9 b, using the prior on $i_*$ derived from \citet{Ahlers+2020} for the fitting model. The black, dashed lines mark the 1-$\sigma$ confidence interval for the values of each parameter, while the blue lines mark the parameter values of the best-fit model solution. See Table \ref{tab:results} for parameter values.
\label{fig:kelt9b_corner_Ahlers}}
\end{figure}

Similar to our analysis of the WASP-33 system discussed above, we apply our model to the KELT-9 b data from the 1st analysis procedure, as shown in Table \ref{tab:b_lambda}, with inflated uncertainties as explained in Section \ref{subsec:system}. 
We use additional orbital constraints from \citet{Ahlers+2020} in our modeling analysis. They observed KELT-9 b's transit light curve to be clearly asymmetric, from which they were able to constrain the value of $i_*$ to be $i_{*,Ahlers}={142}^{+8}_{-7}$~deg\footnote{\citet{Ahlers+2020} defined $i_*$ in their work as $0$~deg when the star is viewed equator on, while in our work $i_*$ is defined as $0$~deg when the ``northern'' pole is directly pointing away from the observer, as shown in Figure \ref{fig:orientation}, which is simply an offset by $90$~deg. We thus translated their published measurement of $i_{*}={52}^{+8}_{-7}$~deg according with our definition in this paper.}. We thus performed our model fit using $i_{*,Ahlers}$ as prior. From the posterior we determined 1-$\sigma$ confidence intervals (determined by the 16th, 50th, and 84th percentiles) and best-fit model values of $J_2$, $i_*$, $\Omega_{0}$, and $I_{0}$, which we list in Table \ref{tab:results} (see Figure \ref{fig:kelt9b_corner_Ahlers} for a visualization of the probability distributions).  

\begin{figure}
    \hspace{0.0\linewidth}
    \includegraphics[width=\linewidth]{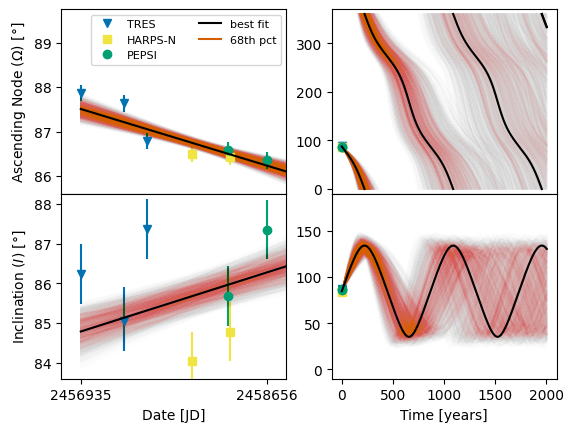}\\
    \includegraphics[width=\linewidth]{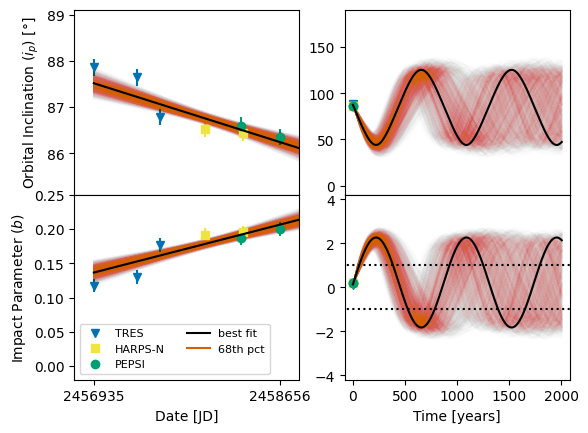}
    \figcaption{Evolution of the ascending node $\Omega$ (first row), inclination $I$ (second row), the planet's orbital inclination versus the observer's line-of-sight $i_p$ (third row), and impact parameter $b$ (fourth row) over time for KELT-9 b, using the prior on $i_*$ derived from \citet{Ahlers+2020} for the fitting model. Dark blue downward triangles and error bars show TRES data, yellow squares and error bars show HARPS-N data, and green dots and error bars show PEPSI data. The black line shows the best-fit model, with red lines showing $2000$ example models with model likelihoods within the 1-$\sigma$ confidence interval around the best-fit model. The precession period for $\Omega$ is approximately $890^{+200}_{-140}$ years. 
\label{fig:kelt9b_precession_Ahlers}}
\end{figure}

Figure \ref{fig:kelt9b_precession_Ahlers} shows the precession evolution based on these fit values, with a precession period of roughly $890^{+200}_{-140}$ years for the ascending node. Like WASP-33 b, KELT-9 b will eventually precess such that it will no longer transit its star from the prespective of an Earth-bound observer, which we estimate will occur around the year $2074^{+12}_{-10}$~CE. KELT-9 b will reappear as a transiting planet approximately $300\pm100$ years afterwards.

We also fit the data with no priors. The resulting posterior distributions are completely consistent with those shown in Fig. 6. The model fit using no priors is provided in the Appendix for comparison.

\begin{table*}
	\centering
	\caption{Modeling Results for the Precessions of WASP-33 b and KELT-9 b. Given are the determined values of the stellar quadrupole moment, $J_2$, the angle of the stellar spin axis versus the line-of-sight, $i_*$, the longitude of the ascending node, $\Omega_0$ and the inclination, $I_0$, at the time of the initial observations, the nodal precession period, and the estimated year when transits will cease. We also estimate the linear ephemerides for the observable parameters $b$ and $\lambda$, with the starting points being the times of the first observations listed in Table \ref{tab:b_lambda} for each system. We give both the 68th-percentile confidence interval for each parameter, as well as the best-fit solutions. Where possible, we compare our results to previous results from the literature.}
	\label{tab:results}
	\begin{tabular}{p{0.2\textwidth}|P{0.15\textwidth}P{0.15\textwidth}|P{0.35\textwidth}} 
	    \multicolumn{4}{c}{WASP-33} \\
		\hline
		Parameter & 68th-percentile confidence interval & Best-fit Solution & Literature Values (References)\\
		\hline
		$J_2$ & $(6.3^{+1.2}_{-0.8})\times10^{-5}$ & $5.92\times10^{-5}$ & $(9.14\pm0.51)\times10^{-5}$ \citep[][]{watanabe20},\newline $(6.73\pm0.22)\times10^{-5}$ \citep[][]{Borsa+2021} \\
		$i_*$ [$\degree$] & $85^{+20}_{-19}$ & $84.63$ & $96^{+10}_{-14}$ \citep[][]{watanabe20},\newline $90.11\pm0.12$ \citep[][]{Borsa+2021} \\
		$\Omega_0$ [$\degree$] & $87.4^{+0.2}_{-0.2}$ & $87.34$ & \\
		$I_0$ [$\degree$] & $109.97^{+0.63}_{-0.64}$ & $109.91$ & \\
		Precession Period [years] & $1460^{+170}_{-130}$ & $1493$ & $\sim 840$ \citep[][]{watanabe20},\newline $1108\pm19$ \citep[][]{Borsa+2021} \\
		Predicted year of last transit [CE] & $2090^{+17}_{-10}$ & $2090.3$ & $ \sim2068$ \citep[][]{Borsa+2021} \\
		$b_0$ & $0.16\pm0.01$ & $0.1609$ &  \\
		$db/dt$ [$years^{-1}$] & $-0.0143\pm0.0013$ & $-0.0145$ &  \\
		$\lambda_0$ [$\degree$] & $-110.0\pm0.5$ & $-109.931$ &  \\
		$d\lambda/dt$ [$\degree years^{-1}$] & $+0.02\pm0.07$ & $+0.019$ & \\
		\hline
		\multicolumn{4}{c}{KELT-9} \\
		\hline
		$J_2$ & $(3.26^{+0.93}_{-0.80})\times10^{-4}$ & $3.38\times10^{-4}$ & $5.6\times10^{-5}<J_2<2.5\times10^{-4}$ \citep[][, estimated]{Ahlers+2020} \\
		$i_*$ [$\degree$] & $139^{+7}_{-7}$ & $139.35$ & $142^{+8}_{-7}$ \citep[][]{Ahlers+2020} \\
		$\Omega_0$ [$\degree$] & $87.47^{+0.18}_{-0.19}$ & $87.51$ & \\
		$I_0$ [$\degree$] & $84.83^{+0.41}_{-0.43}$ & $84.78$ & \\
		Precession Period [years] & $890^{+200}_{-140}$ & $870$ & \textbf{First detection by this work}\\
		Predicted year of last transit [CE] & $2074^{+12}_{-10}$ & $2074.3$ & \\
		$b_0$ & $0.138\pm0.008$ & $0.1366$ &  \\
		$db/dt$ [$years^{-1}$] & $-0.0144\pm0.0025$ & $0.0148$ &  \\
		$\lambda_0$ [$\degree$] & $-84.8\pm0.3$ & $-84.784$ &  \\
		$d\lambda/dt$ [$\degree years^{-1}$] & $-0.31\pm0.08$ & $-0.315$ & \\
		\hline
	\end{tabular}

\end{table*}

\section{Discussion}\label{sec:disc}


\subsection{Improvements on Previous Results and New Findings}\label{subsec:improvements}

One of the goals of this study was to improve upon the established results for the orbital parameters of WASP-33 b and KELT-9 b by previous works \citep[e.g.,][]{johnson15, Gaudi+2017, watanabe20}, given the longer baseline for observations and making use of the measured nodal precession of the planets' orbits. As such, we confirm for WASP-33 b that $i_*$ is close to $90$~deg, and determine a best-fit value for $J_2$ significantly lower compared to previous works with shorter baselines \citep[e.g.,][]{watanabe20}, and consistent with recent work with similar extended baseline coverage \citep{Borsa+2021} (see also Table \ref{tab:results}). 

For KELT-9 b our work presents the first clear detection of the precession of KELT-9 b. We determine a very large $J_2$ value for KELT-9. We caution that the measurement of $i_*$ was complicated due to the variety of observational instruments used for observations. By using the prior based on previous work by \citet{Ahlers+2020}, we confirm that its best-fit value must be close to $140$~deg. Furthermore,  The measured precession rate of approximately $-0.3$~deg per year is consistent with previous predictions by \citet{Ahlers+2020}.

A crucial contribution of our study to the understanding of WASP-33 b and KELT-9 b is in the exploration of both planets' tidal evolution, which we explore more in the following section.

\subsection{Tidal and Orbital Evolution}\label{subsec:tides}

Our observations have clearly shown the tidal influence on the precession of the ascending nodes of the orbits for both WASP-33 b and KELT-9 b, with best-fit values for $J_{2}$, the stellar gravitational quadrupole moments, of $5.92\times10^{-5}$ and $3.38\times10^{-4}$, respectively. These values are in line with expected $J_{2}$ values for fast-spinning stars, which create their quadrupole moments via their oblateness as a response to their rapid rotation \citep[e.g.,][]{Kraft1967,Ward+1976}. The interplay of rotation and internal mass-redistribution is a highly complex phenomenon that depends on factors such as the size of a star's radiative and convective regions, differential rotation rates, magnetic field strengths, and many more. Based on our best-fit parameter results derived from the orbital precession, and defining the stellar oblateness $f$ via the flattening formula \citep{MD00},\begin{equation}
    f=\frac{\Omega_S^2 R_*^3}{2 G M_*}+\frac{3}{2}J_2,
\end{equation}\label{eq:flattening} with $\Omega_S$ being the stellar angular spin frequency, we estimate the stellar oblatenesses of WASP-33 and KELT-9 to be approximately $0.0185\pm0.0025$ and $0.07\pm0.01$, respectively. Our estimate for KELT-9 is consistent with the previous estimate by \citet{Ahlers+2020} of $0.087\pm0.017$.

In our further analysis we are mostly interested in the tidal interactions between the stars and their planets, in particular in terms of the planets' past and future dynamical evolution. Our data clearly shows very strong spin-orbit misalignments. This puts limits on the strengths of the tidal dissipation within the stars that naturally would work towards realignment. 

To estimate these tidal strengths, we assume the equilibrium tide model \citep[following equations by][]{Hut,1998EKH}. In this model an orbiting body raises a tidal bulge on its companion that either lags or leads the bulge-raising object by an angle, which is determined by the ratio of the spin period to the orbital period. In such a situation, dissipative forces then act to both synchronize the spin and orbital periods, as well as realign the orbital inclination and spin axis. The nature of these dissipative forces, however, are not well understood for many objects; in particular, the equation of state and internal structure can majorly impact the strengths of such forces. Previous works have shown that a strong distinction exists between mostly radiative versus mostly convective stars \citep[for an overview, see][]{Ogilvie2014}, with radiative stars generally having much weaker dissipative forces. 

We test both the orbits of WASP-33 b and KELT-9 b using the equilibrium tides model. We determine that only the weaker dissipative forces for mostly radiative stars can allow their planets to exist at their current orbital configurations for an extended period of time. This result is not surprising, as both WASP-33 and KELT-9 are rather large and hot stars, and should both have radiative envelopes. Additionally, we estimate that the tidal dissipation forces acting within the planets vastly out-scale (by at least 5 orders of magnitude during peak migration) the forces within the stars. As such, we determine that (1) the planets' spin periods must be synchronized to their orbital periods, (2) the planet's spin and orbital axes are aligned, and (3) the planetary orbits have no residual eccentricities.

Moreover, we can estimate the timescale of migration. We can assume that the planets were originally formed further away from their hosts and migrated inwards via high-eccentricity migration, which can be caused by a scattering event or secular mechanisms such as the Kozai-Lidov effect by currently unseen companions \citep{Kozai,Lidov,Naoz2016}. This scenario is consistent with the observed large spin-orbit misalignment. 

Starting with original orbital semi-major axes from $1$ to $10$~AU, migration and circularization at their currently observed orbital separations would take approximately $30$ to $100$~Myrs for WASP-33 b and $40$ to $120$~Myrs for KELT-9 b, assuming the equilibrium tide model\footnote{We note here that these time scales depend on the assumed value of the viscous timescale $t_V$, which is usually estimated to be on the order of years or decades for stars and gas giants, as well as the exact nature of the eccentricity-inducing mechanism.}. These time scales are not unrealistic given the estimated ages of roughly $100$~Myrs for WASP-33 and $300$~Myrs for KELT-9 \citep[e.g.,][]{Cameron2010, Gaudi+2017}. However, this would imply that migration began very soon after planet formation. 

These migration scenarios would require a significant dissipation of energy by the planets as heat, reaching on the order of $L_p\sim10^{30}$~ergs/second during the fastest migration episodes, or $\sim0.1\%$ of $L_\odot$, the solar energy output. This energy dissipation is well within realistic levels and would not heat the planets to more than about $2500$~K during peak migration. This calculation ignores additional heat sources and assumes black-body radiation. As the current day-side temperatures of both planets caused by their host stars' incoming irradiation is higher than that, the tidal energy would not significantly alter the planets' chemistry or internal structure. However, if the planets still have any residual internal energy from migration, it could show up as higher-than expected night-side temperatures. If such a discrepancy exists, it would be a telltale sign for migration and allow us to estimate how recently migration must have occurred, with the caveat that winds and other energy transport mechanisms in the planetary atmospheres could dilute such a signal. High night-side temperatures on the order of $2500$~K to $3000$~K have indeed been observed for exoplanets TOI-1431 b/MASCARA-5 b \citep{Addison+2021} and KELT-9 b \citep{Wong+2020}, though in both cases current models explain these temperatures with atmospheric energy transport.

\section{Summary}\label{sec:summary}

Multi-epoch observations of transiting planetary systems can (1) predict the long-term dynamical evolution; and (2) infer stellar and orbital properties that are responsible for the tidal interactions within such systems. We use multi-epoch observations to study two ultra-hot Jupiter systems: WASP-33 b and KELT-9 b. The data spans 11 and 5 years, respectively. Using the Doppler Tomography technique, we adopt an analytical framework to model the high-resolution spectroscopic data sets during the transit. Through the framework, we infer the posterior distribution of the projected obliquity $\lambda$ and impact parameter $b$ over time. The time-varying $\lambda$ and $b$ reveals that both WASP-33 b and KELT-9 b are undergoing orbital precession. 

By applying a tidal interaction model to the time-varying $\lambda$ and $b$ data, we can constrain the $J_2$ values for the two planet host stars (see Table \ref{tab:results}) and predict the dynamical fate of the two systems. We determine the nodal precession periods for the orbits of WASP-33 b and KELT-9 b to be $1460^{+170}_{-130}$ and $890^{+200}_{-140}$ years, respectively. As such, WASP-33 and KELT-9 b are expected to cease transiting their stars around $2090^{+17}_{-10}$~CE and $2074^{+12}_{-10}$~CE, respectively. We furthermore find that either planet can only exist at its current orbit for an extended period of time given weak stellar tidal dissipation consistent with stars with radiative envelopes. The weak tidal dissipation also allows for the observed high stellar spin-orbit misalignments. As such, most tidal dissipation within these systems occurs within the planets, implying planetary spin-orbit alignment and synchronization. If the planets reached their current orbital configurations due to high-eccentricity migration, such migration would have occurred on time scales of a few tens to a hundred Myrs. We estimate that the planets can exist at approximately their current orbital configurations for the remaining main-sequence lifetimes of their host stars, to be eventually engulfed during post-MS inflation.

\section*{Acknowledgements}

A.P.S. acknowledges partial support from a President’s Postdoctoral Scholarship from the Ohio State University and the Ohio Eminent Scholar Endowment. A.P.S. and B.S.G. acknowledge partial support by the Thomas Jefferson Chair Endowment for Discovery and Space Exploration. J.W. acknowledges the academic gift from Two Sigma Investments, LP, which partially support this research. Any opinions, findings, and conclusions or recommendations expressed in this material are those of the authors and do not necessarily reflects the views of Two Sigma Investments, LP. This paper includes data taken at The McDonald Observatory of The University of Texas at Austin, as well as data obtained through the LBT and HARPS-N. The LBT is an international collaboration among institutions in the United States, Italy and Germany. LBT Corporation partners are: The Ohio State University, representing OSU, University of Notre Dame, University of Minnesota and University of Virginia; The University of Arizona on behalf of the Arizona Board of Regents; Istituto Nazionale di Astrofisica, Italy; LBT Beteiligungsgesellschaft, Germany, representing the Max-Planck Society, The Leibniz Institute for Astrophysics Potsdam, and Heidelberg University. The HARPS-N project was funded by the Prodex Program of the Swiss Space Office (SSO), the Harvard University Origin of Life Initiative (HUOLI), the Scottish Universities Physics Alliance (SUPA), the University of Geneva, the Smithsonian Astrophysical Observatory (SAO), and the Italian National Astrophysical Institute (INAF), University of St. Andrews, Queens University Belfast, and University of Edinburgh.

\appendix

\section{Precession Model Fit for KELT-9 b without priors}\label{sec:app}

\begin{figure}
    \centering
    \hspace{0.0\linewidth}
    \includegraphics[width=0.5\linewidth]{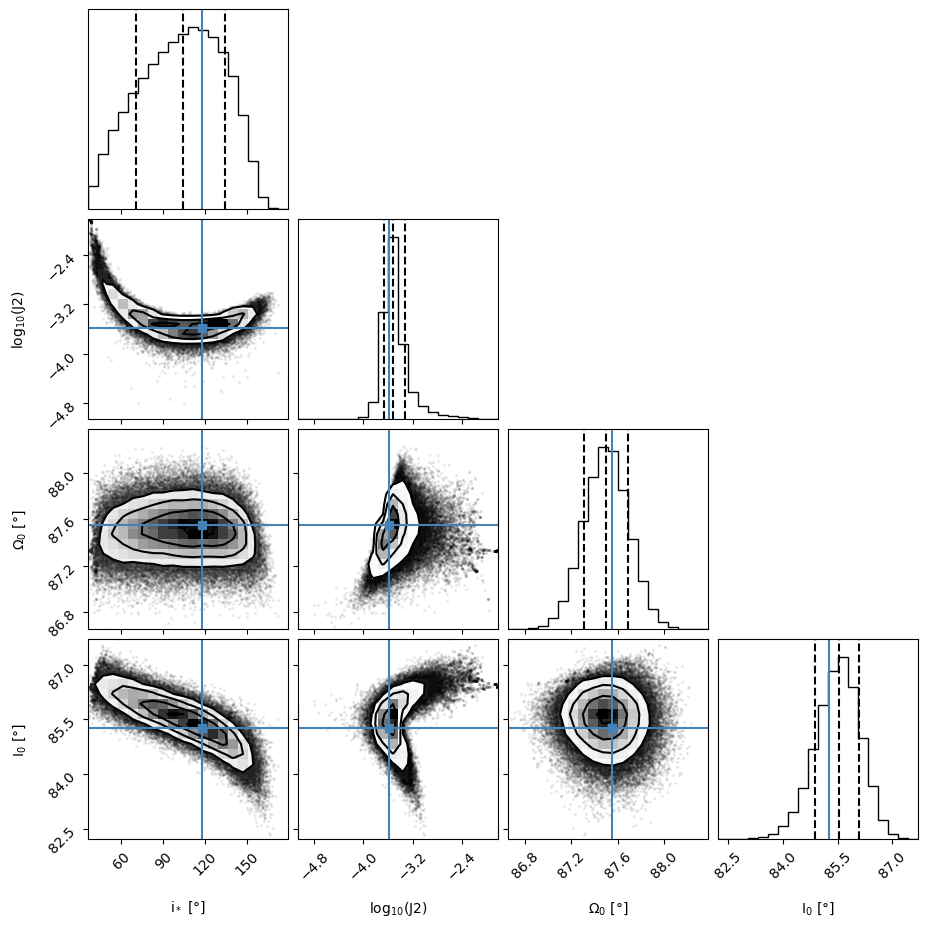}
    \figcaption{Corner plot of MCMC-determined values for $i_*$, $log_{10}(J_2)$, $\Omega_0$, and $I_0$ of KELT-9 b, using no priors for the fitting model. The black, dashed lines mark the 1-$\sigma$ confidence interval for the values of each parameter, while the blue lines mark the parameter values of the best-fit model solution. See text of Section \ref{sec:app} for parameter values.
\label{fig:kelt9b_corner_noprior}}
\end{figure}

\begin{figure}
    \hspace{0.0\linewidth}
    \centering
    \includegraphics[width=0.5\linewidth]{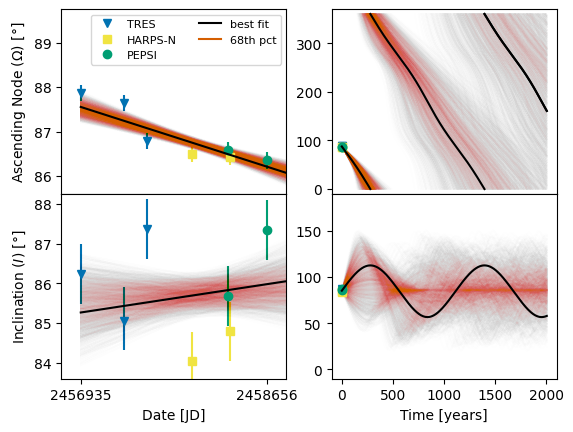} 
    \includegraphics[width=0.5\linewidth]{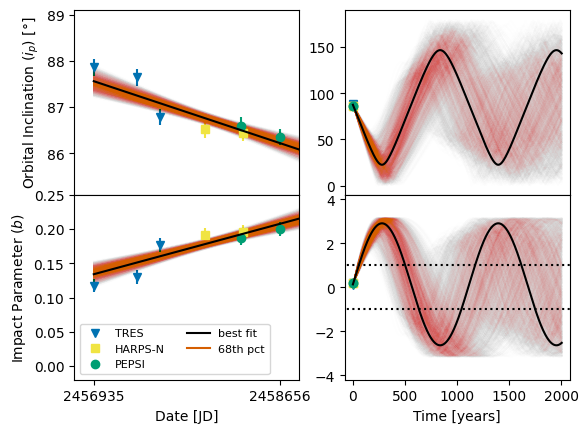}
    \figcaption{Evolution of the ascending node $\Omega$ (first row), inclination $I$ (second row), the planet's orbital inclination versus the observer's line-of-sight $i_p$ (third row), and impact parameter $b$ (fourth row) over time for KELT-9 b, using no priors for the fitting model. Dark blue downward triangles and error bars show TRES data, yellow squares and error bars show HARPS-N data, and green dots and error bars show PEPSI data. The black line shows the best-fit model, with red lines showing $2000$ example models with model likelihoods within the 1-$\sigma$ confidence interval around the best-fit model.
\label{fig:kelt9b_precession_noprior}}
\end{figure}

As mentioned in Section \ref{subsec:prec_KELT9}, our observations for KELT-9 b most likely contained some unknown uncertainties that made fitting the model to the data without additional priors difficult. Here, we provide our model fit to the data using no priors as a comparison to our cited results in Section \ref{subsec:prec_KELT9} using the \citet{Ahlers+2020} prior. Using no priors for the MCMC results in 1-$\sigma$ confidence intervals for the values of the tidal parameter, $J_2=(3^{+1.8}_{-0.8})\times10^{-4}$, of the stellar spin axis angle versus the line-of-sight, $i_*={104}^{+30}_{-34}$~deg, of the initial longitude of the ascending node, $\Omega_{0}={87.50}^{+0.19}_{-0.19}$~deg, and of the initial inclination, $I_{0}={85.53}^{+0.57}_{-0.64}$~deg (see Figure \ref{fig:kelt9b_corner_noprior}). These values correspond to the precession evolution shown in Figure \ref{fig:kelt9b_precession_noprior}. However, the large uncertainty in the value of $i_*$ highlights the limited information content of the data. In particular, the scatter in the observed inclination ($I$) values (see Figure \ref{fig:kelt9b_precession_noprior}, second row, left frame), on the same order as the uncertainties, makes finding an appropriate model fit challenging.
 
\bibliographystyle{aasjournal}
\bibliography{Kozai2} 

\end{CJK*}

\end{document}